\let\DraftKernelLabel\label
\documentclass[
 reprint,
 amsmath,
 pre,
 amssymb,
 footinbib,
 superscriptaddress,
 aps,
 longbibliography,
]{revtex4-1}
\usepackage{graphicx}
\usepackage{xcolor}
\usepackage{array}

\newcommand{\namedref}[2]{\hyperref[#2]{#1~\ref*{#2}}}
\newcommand{\secref}[1]{\namedref{Section}{#1}}
\newcommand{\appref}[1]{\namedref{Appendix}{#1}}
\newcommand{\tabref}[1]{\namedref{Table}{#1}}
\newcommand{\figref}[1]{\namedref{Figure}{#1}}

\newcommand{\tr}{\operatorname{tr}}
\newcommand{\nn}{\nonumber}
\newcommand{\la}[1]{\label{#1}}
\def\<{\langle}
\def\>{\rangle}
\def\[{\left[}
\def\]{\right]}
\newcommand{\beq}{\begin{equation}}
\newcommand{\eeq}{\end{equation}}
\newcommand{\beqa}{\begin{eqnarray}}
\newcommand{\eeqa}{\end{eqnarray}}

\makeatletter
\@ifl@t@r\fmtversion{2023-06-01}{\let\label\DraftKernelLabel\let\ltx@label\label}{}
\makeatother
\usepackage[hidelinks]{hyperref}
\hypersetup{
     colorlinks,
     linkcolor={red!60!black},
     citecolor={blue!70!black},
     urlcolor={blue!80!black}
 }

\DeclareMathSizes{11}{11}{7.7}{5}

\makeatletter
  {\endminipage\endgroup\par\addvspace{\intextsep}}
\newenvironment{placedtable}{\par\addvspace{\intextsep}\noindent
  \begingroup\minipage{\columnwidth}\def\@captype{table}}%
  {\endminipage\endgroup\par\addvspace{\intextsep}}
\makeatother

\begin{document}


\title{\texorpdfstring{Completeness and the Flow I:\\
Correlation functions in $\mathcal{N} = 4$ SYM}{Completeness and the Flow I: Correlation functions in N = 4 SYM}}


\author{Gabriel Lefundes}
\affiliation{
IFT-UNESP, ICTP-SAIFR,
Rua Dr Bento Teobaldo Ferraz 271,
01140-070 SP, Brazil
}
\affiliation{
Université Paris-Saclay, CNRS, CEA,
Institut de Physique Théorique,
91191 Gif-sur-Yvette, France
}

\author{Pedro Vieira}
\affiliation{
IFT-UNESP, ICTP-SAIFR,
Rua Dr Bento Teobaldo Ferraz 271,
01140-070 SP, Brazil
}
\affiliation{
Perimeter Institute for Theoretical Physics,
31 Caroline St N, Waterloo,
Ontario N2L 2Y5, Canada
}

             
\begin{abstract}
Completeness in quantum mechanics equates a sum over all states to the identity. Other sums over states appear in amplitudes, partition functions, and correlation functions. Can these be obtained by suitably dressing the completeness relation? Can this dressing be generated by a differential flow interpolating between the identity and the physical quantity of interest? Can such flows evaluate the sums without computing their individual terms? We explore these questions -- and answer them all in the affirmative -- for the simplest connected four-point functions in planar $\mathcal{N}=4$ SYM at tree level. We discuss several extensions including some opportunities to extend our construction to finite coupling on the basis of recent developments. 
\end{abstract}

\maketitle
\section{Introduction} 
We love completeness 
\beq
\mathbb{I}=\sum_\psi  \frac{|\psi\rangle \langle \psi|}{\langle\psi |\psi\rangle} \,. \label{completeness} 
\eeq
Here we suggest exploiting completeness to compute  more interesting sums in quantum theories. The basic idea will be to construct interesting difference-differential operators that act on this trivial relation transforming it into a nice physical observable. 

While the strategy is much more general, we will illustrate the idea in planar $\mathcal{N}=4$ SYM at weak coupling for the four-point function $G(z,\bar z)$ of the chiral primary operators in the limit $\bar z\to 0$ where the correlator $G(z,\bar{z})\rightarrow G(z)$ is dominated by the leading-twist operators in the theory which span a nice closed subsector known as the $SL(2)$ sector~\cite{Beisert:2003yb,Beisert:2003jj,Braun:2003rp,Belitsky:2004cz,Belitsky:2004sc}.

Explicitly, the OPE decomposition reads \cite{Dolan:2000ut}
\beqa
&&\!\!\!\!\!\!\!G(z)=\sum_{\psi }\frac{\<12|\psi \>\<\psi|34 \>}{\langle\psi |\psi\rangle} z^S {}_{2}F_1(\ell\!+\!S,L\!-\!\ell\!+\!S;L\!+\!2S; z)  \nn\\
&&\!\!\!\!\!\!\! \psi= \text{zero-momentum primaries, with spin $S$}  \label{untwisted}
\eeqa
where we sum over single-trace primaries $\psi$. The descendants are captured by the conformal block given by the hypergeometric function. The prefactor is given by the  (squared) tree-level structure constants which can be computed from integrability \cite{Escobedo:2010xs,Vieira:2013wya,Sobko:2013ema}. Each state $\psi$ is a quantum state of an integrable $SL(2)$ nearest neighbor closed spin chain and the single-trace condition from the planar limit tells us to stick to spin chain states with zero momentum~\cite{Minahan:2002ve,Beisert:2003yb}. The precise definition of the correlator -- which will clarify the meaning of the bridge length $\ell$ and total length $L$ -- will be given soon. 

Our approach here will be very much along the lines of the so-called separation of variables (SoV) approach \cite{Sklyanin:1995bm,Derkachov:2001yn,Derkachov:2002tf}. 
We will discuss the interplay with the more established hexagon approach \cite{Basso:2015zoa,Basso:2025mca} below. 

The sum \eqref{untwisted} looks considerably more involved than \eqref{completeness}; we only sum over the subset of closed spin chain states that have zero momentum (single traces) and which are killed by $\mathsf{K}$ (primaries) and we weight each term in the sum by a relatively complicated conformal block. We can relax the zero-momentum condition by projecting our correlator onto zero-momentum states through a simple sum. Finally, we relax the restriction to primaries by introducing a small twist $\phi$ -- a procedure which breaks conformal symmetry and renders the descendants independent from the primaries~\cite{Deguchi_2001,Beisert:2005if,Kazakov:2015efa,Cavaglia:2020hdb,Bargheer:2026kon} -- which we set to zero at the end to get
\beqa
&&\!\!\!\!\!\!\!\! G(z)=\lim_{\phi \to 0} \frac{1}{L} \sum_{n=0}^{L-1} \<12|\mathbb{K} | 34\>\,, \,\,\,\,\, \mathbb{K}=
\sum_{
\psi
}
 \frac{|\psi \>\<\psi|}{\langle\psi |\psi\rangle} z^S e^{i P n}  \,. \nn\\
&&\!\!\!\!\!\!\!\! \psi= \text{all states, with momentum $P$ and spin $S$} 
\eeqa
These sums are already quite close to \eqref{completeness}! Indeed, ignore the momentum factor for now and suppose we could find a simple representation of the spin operator. Then we could simply write 
\beq
\mathbb{K}=z^{\hat{S}}  \circ \mathbb{I}
\label{expflow}
\eeq
and declare victory as long as we could efficiently evaluate the action on the right-hand side. Equivalently, we could convert this relation into a \textit{flow} equation 
\beq
(z \partial_z-\hat{S}) \mathbb{K}=0 \,, \label{flow}
\eeq
which we would solve with the prescribed completeness boundary conditions at $z=1$. Instead of painfully computing tons of structure constants with integrability and adding them up \cite{Basso:2017khq} as in \eqref{untwisted}, we simply need to solve a beautiful flow differential equation \eqref{flow}. 
\section{\texorpdfstring{$G(z)$}{G(z)} and a sum of skeletons}

\begin{figure}[t]
    \centering
\includegraphics[width=\linewidth]{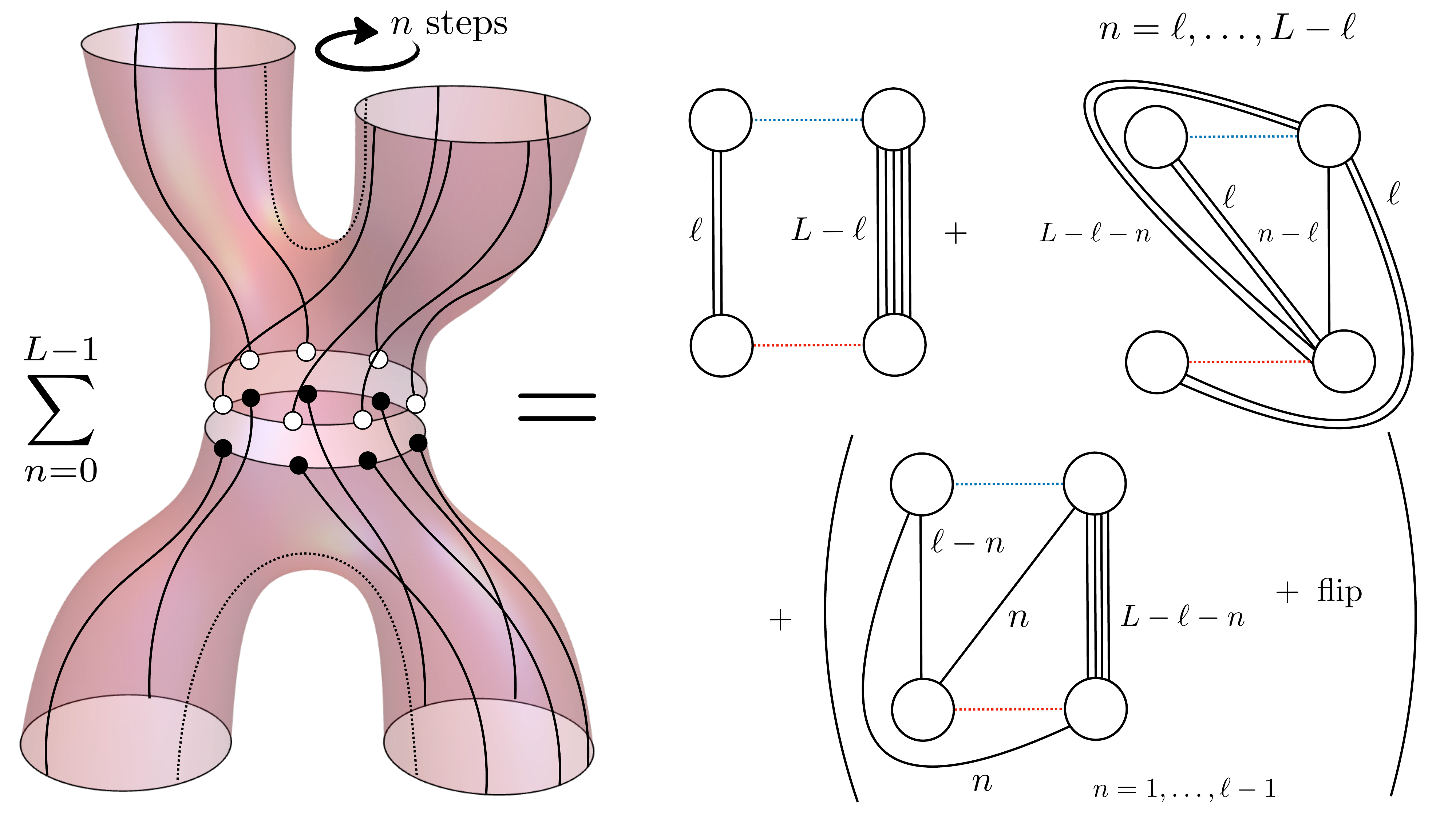}
    \caption{
    Diagrams contributing to the four-point function of the operators $\tr(XZ^\ell)(x_1),\allowbreak \tr(\bar XZ^{L-\ell})(x_2),\allowbreak  \tr(Y\bar Z^\ell)(x_3)$, and $\tr(\bar Y \bar Z^{L-\ell})(x_4)$. (In the figure $\ell=2$ and $L=7$.) We can think of each of the $L$ diagrams as taking the pair of pants at the top and rotating it by $n=0,1,\dots,L-1 $ units before gluing it to the pair of pants at the bottom. On the left we depict $n=1$. Each skeleton graph is given by an OPE sum over \textit{all} states but the sum of rotated configurations when gluing the two pairs of pants projects onto intermediate zero momentum spin chain states. }
    \label{fig4pt}
\end{figure}

Consider the tree-level correlation function in \figref{fig4pt}. The reduced tree-level correlator in the light-cone limit 
\beq
G(z) \equiv \lim_{\bar z\to 0} \[ x_{12}^2 x_{34}^2 x_{13}^{2\ell}x_{24}^{2L-2\ell}\times \<O_1 O_2 O_3 O_4 \> \] \,,
 \eeq
is given by the sum of $L$ diagrams in the figure, that is, 
\beq
G(z)=\sum_{n=0}^{L-1} G_n(z)  = \sum_{n=0}^{L-1} \frac{1}{(1-z)^{\min(n,\ell,L-n)}}\,.  \label{sumSkeletons}
 \eeq
Each of these $L$ terms is called a \textit{skeleton} and we have assumed without loss of generality $\ell\leq L-\ell$.

It is amusing to note that we can OPE decompose not only the full correlator but actually each skeleton on its own. Indeed, we find that we can reproduce each of the $L$ skeletons by an OPE sum over \textit{all} closed spin chain states (not only zero-momentum states) as long as we weight each flowing state with $e^{i P n}$ where $P$ is the momentum of the state. When we put all diagrams together, that sum of exponentials of momentum simply projects onto zero-momentum states as usual:
\beq
\sum_{n=0}^{L-1} e^{i P n}=\frac{e^{i P L}-1}{e^{i P}-1} \, .
\eeq
It vanishes for all physical states of a closed chain since those obey $e^{i P L}=1$, except of course when $P=0$. 
We even checked the one-loop OPE decomposition of each one-loop quantum corrected skeleton, see \appref{OPEskeletonsAp}. The one-loop check is particularly interesting since at one loop only a few skeletons are corrected, meaning that there are remarkable cancellations in all other skeletons such that the one-loop sum rules vanish. 

The split into skeletons is also quite appealing from a hexagonalization point of view: There, the very first step is to split the correlator into these skeletons, build a triangulation for each skeleton and then promote the triangles to full-fledged quantum hexagons with precise quantum sums over flowing states along the lines gluing these triangles~\cite{Fleury:2016ykk,Eden:2016xvg,Coronado:2018ypq}. We suggest that each such skeleton has a nice OPE decomposition on its own. 
We will also show that each skeleton correlator is a solution to a simple differential equation which would be nice to derive directly from hexagons, thus providing a nice bridge between the SoV and hexagon approaches, see also \cite{Jiang:2015lda,Jiang:2016ulr,Bargheer:2026kon}. 

\section{Integrability, SoV, and the twist}
How to find all the spin chain states and compute all their correlation functions with integrability by means of SoV is well known at weak coupling. States are solutions to the $SL(2)$ Baxter equation \cite{Beisert:2003jj,Derkachov:2002tf,Braun:1999te,Belitsky:2004cz} and structure constants at the first loop orders can be computed through SoV as detailed in \cite{Bercini:2022jxo}; see \appref{UntwistedBaxterSoVOPE}.  Here let us present a twisted version of this story. We introduce a twist $\phi$ which we will send to zero at the end~\footnote{The twist introduced here must have some nice geometrical meaning as in \cite{Cavaglia:2018lxi,Cavaglia:2020hdb,Beisert:2005if,Bargheer:2026kon}. It would be nice to pursue that connection further; for now, take it as a convenient trick.}. The twisted Baxter equations take the form
\beq
\hat{\mathcal{B}}_u \circ Q(u) =T(u) Q(u) \label{baxter}
\eeq
where the Baxter operator is given by
\beq
\hat{\mathcal{B}}_u = e^{i \phi} (u+i/2)^L e^{i\partial_u}  +e^{-i \phi} (u-i/2)^L e^{-i\partial_u}\, ,
\label{baxterop}
\eeq
$Q(u)$ is a polynomial of degree $S$, and the transfer matrix 
\beq
T(u)=2\cos(\phi) u^L -\sin(\phi)(2S+L) u^{L-1} + \sum_{k=0}^{L-2} c_k u^k 
\eeq
is a degree $L$ polynomial with its two leading coefficients fixed as written here. Importantly, the spin enters linearly in this expression.  This will be very useful later. (In the untwisted story, the spin enters quadratically in the $u^{L-2}$ term through the $SL(2)$ Casimir combination, see \appref{untwistedAp}.)

Each state is a solution $Q(u)$ of the Baxter equation \footnote{Its spin chain one-loop energy -- which gives the one-loop anomalous dimension of these states -- is proportional to $Q'(i/2)/Q(i/2)-Q'(-i/2)/Q(-i/2)$. We quote this result for completeness only and because the energy is of course the simplest observable to extract out of a quantum state but we will not use it here.}. The momentum of the state is given by~\footnote{With the twist, $e^{iPL}$ is no longer equal to $1$ but rather equal to $e^{-2i\phi S}$. However, that doesn't matter since we want to take $\phi \rightarrow 0$ in the end.} 
\beq
e^{iP}=Q(-i/2)/Q(i/2) \,. \label{PQ}
\eeq
When we introduce a twist we should now sum over all states; there is no longer a split between primaries and descendants since the $SL(2)$ symmetry which relates them is now explicitly broken by the twist. This can be nicely seen for $L=2$. For $L=2$ we can easily verify that there is $1$ solution of \eqref{baxter} for each spin $S$ when $\phi=0$ but $S+1$ solutions when $\phi \neq 0$. As we send $\phi\to 0$, the additional $S$ solutions at each spin become the descendants of the lower spin solutions. (What happens is that those $S$ solutions are polynomials where some of the zeros -- called Bethe roots -- go to infinity as $\phi\to 0$ and roots at infinity in the untwisted theory are nothing but symmetry generators -- see e.g. \cite{Faddeev:1996iy, Deguchi_2001,Beisert:2003yb,Beisert:2005di,Beisert:2005fw} -- in this case $\mathsf{P}$ raising operators of the $SL(2)$ symmetry of the untwisted correlator.) 
We can now sum over all solutions to the twisted Baxter equations and weight those by $z^S$ instead of summing over primaries dressed by a complicated conformal block. 

Finally, to construct the relevant structure constants arising in the OPE sum, we build the SoV scalar product \cite{Derkachov:2002tf,Kazama:2013rya,Cavaglia:2019pow,Bercini:2022jxo}
\beq
\< Q_1,Q_2\>_\ell \equiv \int \, d\mu_\ell(u_1,\dots,u_\ell) \prod_{k=1}^\ell Q_1(u_k)Q_2(u_k) 
\eeq
where we use measures normalized by $\int d\mu_m(\vec u)=1$ (and $d\mu_0=1$). Their explicit expressions are quoted in \appref{UntwistedBaxterSoVOPE}. For example, 
\beq
\mu_1(u)=2\sin\phi\,e^{2\phi u}/(1+e^{2\pi u}) \,. \label{mu1}
\eeq
For the bootstrap of these SoV measures, the study of their properties, the extension of these scalar products to loop level, the orthogonality of Baxter solutions under such scalar products, the connection between such scalar products and other integrable scalar products such as the Gaudin norm and so on, see \cite{Derkachov:2002tf,Cavaglia:2019pow,Gromov:2019wmz,Gromov:2020fwh,Bercini:2022jxo,Bargheer:2025kli,Korepin:1982gg}.

The structure constants can then be built out of these scalar products  as (a minor twisted version of what is) explained in \cite{Bercini:2022jxo} and recalled in \appref{UntwistedBaxterSoVOPE}. 

\section{From \texorpdfstring{$\mathbb{K}$}{K} to the many skeletons}
We can now construct our fundamental object, which we dub \textit{the kernel} $K(u_1,\dots,u_L,v_1,\dots,v_L) $, as
\beq
K \equiv \sum z^S \frac{Q(u_1)\dots Q(u_L)Q(v_1)\dots Q(v_L)}{\<Q,Q\>_L} 
\label{defkernel}
\eeq
where the sum is over all Baxter solutions. Note $K$ is symmetric in all its $2L$ arguments. From this master object we can get $G(z)$ easily as 
\beq
G(z)= \sum_{n=0}^{L-1}  G_n (z)\,,
\eeq
where each term in the sum is given by
\begin{multline}
G_n(z)=\lim_{\phi \to 0}\int d\mu_\ell(u_1,\dots,u_\ell) d\mu_{L-\ell}(u_{\ell+1},\dots,u_L) \, \times\\
\qquad K(u_1,\dots,u_L,+\tfrac{i}{2},\dots,+\tfrac{i}{2},\underbrace{-\tfrac{i}{2},\dots,-\tfrac{i}{2}}_\text{$n$ times}) 
\label{eq:Gn}
\end{multline}
Here $n$ counts the rapidities set to $-i/2$, with the remaining $L-n$ set to $+i/2$. In \appref{properties}, we present some nice features of this object.

Note that each of the $L$ terms in the sum is supposed to precisely reproduce a corresponding skeleton, i.e. one term in \eqref{sumSkeletons}. Indeed, due to \eqref{PQ}, the $Q$-functions evaluated at $v_j = \pm i/2$ precisely generate the $e^{iPn}$ factors mentioned above and 
 the integration with the two measures precisely constructs the two structure constants in the OPE (note that one structure constant is given by $\ell$ integrals and the other is given by $L-\ell$ integrals; this is a (convenient) choice so that in the end we have precisely~$L$ integrations.\footnote{As explained in \cite{Bercini:2022jxo} each structure constant has a nice $\ell \leftrightarrow L-\ell$ symmetry so we could have used other representations with $2\ell$ or $2L-2\ell$ Q's for example; why we prefer to have $L$ $Q$-functions $Q(u_j)$ in our kernel and thus in \eqref{eq:Gn} will be clear in the next section. Another reason for that choice is that there are no further combinatorial factors in \eqref{eq:Gn} dressing the various SoV scalar products as in \cite{Bercini:2022jxo}; this is one of the remarkable features of the twisted expression. All the combinatorics -- in particular the various relative factors giving rise to the conformal blocks -- arise automatically when taking $\phi\to 0$, see also \appref{untwistingAp}. })
All this is only interesting if we can find nice equations for this kernel without explicitly summing over solutions to the Baxter equation (we can; see \secref{secFlow}), if we can then efficiently perform the integrals in \eqref{eq:Gn} (we can; see \secref{IntSec}), 
\textit{and} if we can, moreover, extend it to loops and beyond the $SL(2)$ sector (we hope so; see \secref{Discussion}). 

\section{Spin operator and the flow(s)} \label{secFlow}
Some of the manipulations in this section and the next can be a bit notation heavy; to develop intuition for the various ingredients we strongly recommend also checking out \appref{L1kernelAp} for the very simplest~$L=1$ case.

We now turn our attention to making the flow equation \eqref{flow} precise. The missing ingredient is the spin operator. 

A key observation is that the spin is nothing but a coefficient in the degree $L$ polynomial $T(u)$; because $T(u)$ is a polynomial we can extract any of its coefficients from its leading coefficient and its values at $L$ distinct points $u_1,\dots,u_L$. A Baxter operator acting on a $Q$-function produces precisely one such transfer matrix factor (\ref{baxter}) -- see also \cite{Derkachov:2002tf,Gromov:2022waj} -- so with $L$ $Q$-functions we can easily extract the spin operator. We conclude that 
\beq
\hat{S}_{\vec u}\prod_{j=1}^LQ(u_j)=S\prod_{j=1}^LQ(u_j),
\eeq
with the spin operator 
\beq
\hat{S}_{\vec{u}}=\frac{e^{i \phi}\bar D_{\vec{u}}-e^{-i \phi}D_{\vec{u}}}{2i\sin\phi}.
\label{spinOp}
\eeq
Here, 
\beq
D_{\vec{u}} = i \sum_{j=1}^L \frac{(u_j-i/2)^L}{\prod_{k\neq j} (u_j-u_k)}\left(e^{- i \partial_{u_j}}-1\right)
\label{Ddef}
\eeq 
and $\bar{D}$ is obtained from $D$ by $i\rightarrow-i$. 

We are now ready to get our flow equation: it follows from the definition of the kernel, equation \eqref{defkernel}, that the action of $z \partial_z$ has the same result as the action of the operator $\hat{S}_{\vec{u}}$. Indeed, since the spin operator acts on $L$ $Q$-functions, while the kernel contains $2L$ of them, we get one flow equation for each subset $I$ of the $2L$ rapidities with length $L$. Denoting its complementary set by $\bar{I}$, 
\beq
(z\partial_z-\hat{S}_I)K(I,\bar{I})=0. \label{flowEqRig}
\eeq
This is the sought-after precise version of \eqref{flow}. If an LLM were typing this draft, it would certainly put this equation inside a nice box. 

Similarly, the precise version of \eqref{expflow} is
\beq
z^{\hat{S}_{I}} f(I) = \int d\mu_L(\bar{I}) K(I,\bar{I}) f(\bar{I})
\label{expflow2}
\eeq
where $f(I)$ is a test function completely symmetric in its arguments. We emphasize this representation encodes \textit{both} the flow equations (when $z\neq 1$) \textit{and} completeness (for $z=1$). 

In fact, the kernel obeys not one but \textit{two} sets of nice flow equations given by 
\beqa
&&\!\!\!\!\!\!\!\!\!\!\!\! \!\!\! D_IK \!=\!\left[z\partial_z\!-\frac{z e^{+i \phi}\sin\phi}{1-z}
(\partial_\phi+\!2C_{\bar{I}}+\!L\cot\phi)\right]\!K \nn\,,\\
&&\!\!\!\!\!\!\! \!\!\!\!\! \!\!\! \bar{D}_IK \!=\!\left[z\partial_z\!-\frac{z e^{-i \phi}\sin\phi}{1-z}
(\partial_\phi+\!2C_{\bar{I}}+\!L\cot\phi)\right]\!K \label{eq:Dequation2} \,,
\eeqa
where $C_{\bar{I}}$ is the sum of all rapidities in the set $\bar{I}$. As a first consistency check note that multiplying the first/second flow equation in \eqref{eq:Dequation2} by $e^{-i \phi}$/$e^{+i \phi}$ and subtracting them nicely leads to the original flow equation \eqref{flowEqRig}. 

These two flow equations are derived very straightforwardly once we observe that the operators $D$ and $\bar{D}$ form a nice closed algebra, $[D,\bar{D}] = D-\bar{D}$. Using these commutation relations, we can easily derive the equation 
\beq
D_{I} z^{\hat{S}_{I}} = \left(z \partial_z - \frac{z e^{i \phi} \sin \phi}{1-z} \partial_\phi \right) z^{\hat{S}_{I}}
\eeq
and the conjugate one for $\bar{D}$. These, in turn,  immediately lead to the flow equations \eqref{eq:Dequation2} once we apply these operators to the exponentiated flow relation \eqref{expflow2}. The last terms in the square brackets appear due to the action of the $\phi$ derivative on the measure $d\mu_L(\bar{I})$. 

We saw that the (weighted) sum of the two flow equations in \eqref{eq:Dequation2} leads to the original flow equation \eqref{flowEqRig}; what else do these equations yield? Their compatibility leads to what we call \textit{covariance}: the statement that apart from a simple factor, the dependence on the cross-ratio and the twist only shows up in a very precise combination 
\beq
\chi = -4 z \sin^2 \phi/(1-z)^2\,,
\label{eq:chi}
\eeq
namely
\beq
K=
\frac{(1-ze^{2i\phi})^{i(u_1+\dots+v_L)}}{(1-z)^{L+i(u_1+\dots+v_L)}}
\!\times\! H(u_1,\dots,v_L|\chi)\,.
\label{eq:covariance}
\eeq
In \appref{L1kernelAp} and \appref{L2kernelAp} we present explicit flow solutions for these kernels for $L=1$ and $L=2$. 

\section{Integrating the flow} \label{IntSec}

To get the physical correlators, we want to subject the kernels to two important operations: we set the auxiliary variables $v_j$ to $\pm \frac i2$ and we perform the integrations over the rapidities $u_j$ -- see \eqref{eq:Gn}. Which operation we carry out first is a matter of choice and different orders have different pluses and minuses. Remarkably, we can do both without ever explicitly solving the flow equations but instead by exploiting several of its nice properties.

For example, by definition \eqref{Ddef}, $D$ and $\bar{D}$ have the useful property that at $+i/2$ or $-i/2$ they reduce to lower~$L$ operators acting on a smaller set:
\beq
D_{I\cup\{+i/2\}}\longrightarrow D_I,\qquad
\bar D_{I\cup\{-i/2\}}\longrightarrow\bar D_I.
\label{eq:reduction}
\eeq
This property will allow us to unveil remarkable simplifications when we set the $v$ rapidities to $\pm i/2$. On the other hand, if instead we integrate over the rapidities first we encounter another remarkable simplification: the twist dependence totally drops out, even before specifying any values for the $v_j$!
\beq
\partial_\phi \int d\mu_\ell(U)\,d\mu_{L-\ell}(W)\,K(U,W,v_1,\dots,v_L)=0\,,
\eeq
where $U \equiv (u_1,\dots,u_\ell)$ and $W \equiv (u_{\ell+1},\dots,u_L)$. This follows easily from the  relations derived in the previous section together with the explicit form of the SoV measures, see 
\appref{PhiIndAp}.

With equations for the kernel established, we will now set the auxiliary variables $v_j$ to $\pm \frac i2$ and perform the integrations over the rapidities $u_j$ as prescribed in \eqref{eq:Gn}. The computation depends on the value of $n$, the number of $-i/2$'s. The case $\ell\leq n\leq L-\ell$ is the simplest. In this case we can construct two nice sets of rapidities: 
\beq
I_{\pm} =W\cup\{\underbrace{\pm \tfrac{i}{2},\dots,\pm \tfrac{i}{2}}_\text{$\ell$ times}\} \,,
\eeq
each with $L$ elements, and write
\beq
\frac{e^{i \phi} \bar{D}_{I_-} - e^{-i \phi} D_{I_+}}{2 i \sin \phi} K = \hat{S}_W K = \left(z \partial_z - \frac{\ell z}{1-z}\right) K \,. \nn
\eeq
In the first equality we have used the fact that the operators $D_{I_+}$ and $\bar{D}_{I_-}$ reduce to $D_{W}$ and $\bar{D}_{W}$ via \eqref{eq:reduction}. Then we recognize what's left as the spin operator for a smaller length $L-\ell$, due to \eqref{spinOp}. On the other hand, we can compute the LHS using the equations \eqref{eq:Dequation2}. In the chosen combination, the derivatives in $\phi$ cancel and we are left with the RHS.
To get an equation for the correlator $G_n$ itself, we need to integrate the last equality against the measures $d\mu_\ell$ and $d\mu_{L-\ell}$:
\begin{multline}
    \left(z \partial_z - \frac{\ell z}{1-z}\right) G_n \!=\! 
    \int \!d\mu_\ell(U) d\mu_{L-\ell}(W)  
\hat{S}_W\bigl[K\bigr]. \nn
\end{multline}
Integration is now trivial, since we can move the spin operator past the measure $d\mu_{L-\ell}$ by shifting contours such that it now acts on the constant function $1$, yielding zero. Using $G_n(0)=1$ -- true for any $n$ by definition -- we therefore conclude that
\beq
\left[(1-z)\partial_z-\ell\right]G_n=0 \,\,\,\, \Longrightarrow\,\,\,\,  G_n=(1-z)^{-\ell}. \label{eq1}
\eeq
The complementary cases, where $n<\ell $ or $n> L - \ell$, are slightly different. They involve a slightly more clever choice of rapidities and a tiny bit of extra algebra -- see \appref{OtherCase}. The outcome is an almost identical very simple differential equation with an equally simple solution. For example, for $n<\ell$,
\beq
\left[(1-z)\partial_z-n\right]G_n=0 \,\,\,\, \Longrightarrow\,\,\,\,G_n= (1-z)^{-n} \,. \label{eq2}
\eeq
This concludes the derivation of \eqref{sumSkeletons}. As promised, we ended up not solving a single Baxter equation nor performing any painful sum over intermediate states.  

\section{Discussion and Musings}  \label{Discussion}
Integrability allows us to compute the spectrum of any single-trace operator in planar $\mathcal{N}=4$ SYM at any coupling \cite{Beisert:2010jr,Gromov:2009tv,Gromov:2009bc,Bombardelli:2009ns,Arutyunov:2009ur,Gromov:2013pga,Gromov:2014caa} and, to some extent, their three-point correlation functions as well \cite{Basso:2015zoa,Basso:2015eqa,Basso:2017khq,Basso:2022nny,Basso:2025mca,Ekhammar:2026sli}. In a CFT, these numbers constitute the CFT data. What distinguishes the dynamics of different CFTs is of course the behavior of higher-point correlation functions which are by no means kinematical. Such objects are made out not of individual elements of the CFT data but rather of nice sums over these data. Here we ask whether these sums might be simpler to evaluate directly rather than by adding things up individually. 

The master kernel \eqref{defkernel} allows us to do precisely this. We managed to transform completeness relation sums -- present in any quantum theory -- into simple differential equations for the simplest possible higher point correlators, see e.g. \eqref{eq1} and \eqref{eq2}. 

We look forward to generalizing these equations and unveiling their physical meaning. Can we go beyond the leading twist limit $\bar z \to 0$ which projected us into the $SL(2)$ closed sector of the theory? Can we extend the computation of the kernels and the differential equations to loop level and even finite coupling? 

A few years ago \cite{Basso:2022nny} Basso, Georgoudis and Klemenchuk Sueiro put forward a beautiful expression for the full finite coupling resummation  structure of the hexagon sums precisely for the $SL(2)$ three-point functions required here. And last week~\cite{Ekhammar:2026sli} massaged their resummation into a very inspiring form strongly reminiscent of a sort of all loop SoV representation. The all loop structure constant factorizes into three blocks as $C=(\texttt{normalization factor})\times \mathcal{A} \times \mathcal{B}_p$ where $\mathcal{A}$ and $\mathcal{B}_p$ resum all adjacent and bottom wrapping contributions respectively \cite{Basso:2022nny}. For twist-two operators these two wrapping contributions are actually related to each other as 
\beq
\mathcal{A} = (S-1) g^2 |\hat{\alpha}| \frac{\Gamma((\Delta+S-4)/2)}{\Gamma((S+4-\Delta)/2)} \mathcal{B}_2 \label{B2}
\eeq
which is a small extension to general spin $S$ of the relation in~\cite{Ekhammar:2026sli} derived for the Konishi operator ($S=2$) and where $\hat{\alpha}$ is given by the large-$u$ asymptotics of the QSC $Q$-functions, as described in their paper. We checked this conjecture to five-loop order using AI. What is more, we can actually make this relation even nicer if we introduce the same twist $\phi$ we introduced above:
\beq
\mathcal{A} =  \lim_{\phi \to 0} (-2i\phi)^{\Delta-2} \, \mathcal{B}_0 \label{B0}
\eeq
so that the untwisting procedure actually nicely yields all the gamma-function pre-factors automatically! In \appref{app:finite} we explain how (\ref{B0}) leads to (\ref{B2}) and explain how this $\mathcal{B}_0$ object precisely makes contact with our SoV integrals in (\ref{eq:Gn}) at weak coupling. The normalization factor has not yet been cast as a nice SoV representation but this might not be needed actually. In all the derivations above we never used the precise representation of the norm as an SoV scalar product; all we needed was the completeness relation. 
If true, for the leading twist contribution to the simplest $20'$ correlator say -- i.e. to compute $G(z)$ at finite coupling -- we would ``only" need to sum a product of four $\mathcal{B}$ factors (namely $\mathcal{B}_0\mathcal{B}_0\mathcal{B}_2 \mathcal{B}_2$) for all twisted twist-two states in the OPE expansion. Are the ideas unveiled here useful for directly performing this sum? Can we get rid of the Baxter functions altogether and unveil some extension of the differential-integral equations derived by Belitsky and Korchemsky in \cite{Belitsky:2019fan,Belitsky:2020qir,Belitsky:2020qrm} for the octagon correlators \cite{Coronado:2018ypq,Coronado:2018cxj}?, see also \cite{Kostov:2019auq,Kostov:2019stn}. We are currently exploring this promising possibility.

Maybe, at the end, everything will reduce to a nice big determinant. 

{\begin{center}{\textbf{ACKNOWLEDGMENTS}} \end{center}}
We thank Gleb Arutyunov, Carlos Bercini, Yagmur Bulbul, Frank Coronado, Gwenaël Ferrando, Kolya Gromov, Alexandre Homrich, Shota Komatsu, Gregory Korchemsky, Vincent Pasquier, Paul Ryan and Didina Serban for enlightening discussions. We acknowledge the discussions and useful feedback from the participants of IGST 2026 at DESY when this work was first presented. 
GL is supported by the Conselho Nacional de Desenvolvimento Científico e Tecnológico (CNPq) - Grant 141597/2025-9. GL acknowledges the Perimeter Institute for hospitality and financial support during the period this work was finished. PV is supported in part by Discovery Grants from the Natural Sciences and Engineering Research Council of Canada.
Research at the Perimeter Institute is supported by the Government of Canada through the Department of Innovation, Science and Industry Canada, and by the Province of Ontario through the Ministry of Colleges and Universities. \\

We used LLMs to assist in small algebra computations and derivation cross-checks as well as to check the relation (\ref{B2}) perturbatively. We independently checked all the results of this draft by hand or Mathematica and we did not use any AI to write the draft. 
\appendix
\setcounter{secnumdepth}{2} 

\section{Four-point functions and skeletons}
\label{OPEskeletonsAp}
\subsection{OPE review}
We want to study the correlator
\beq
\langle \tr(XZ^\ell)(x_1) \tr(\bar XZ^{L-\ell})(x_2)\tr(Y\bar Z^\ell)(x_3)\tr(\bar Y \bar Z^{L-\ell})(x_4)\rangle\,. \nn
\eeq
Its OPE decomposition in the limit $\bar{z}\rightarrow0$ is~\cite{Dolan:2000ut}, after extracting a factor $x_{12}^{-2} x_{34}^{-2} x_{13}^{-2\ell} x_{24}^{-2(L-\ell)}$, 
\beq
G(z) = \sum_{\substack{\mathcal{O}\\ \text{zero-mom}}} \mathcal{F}_\mathcal{O}\times C_{ \mathcal{O}}^2 
\eeq
The cross-ratios $z$ and $\bar{z}$ are defined by
\beq
z \bar{z} = \frac{x_{12}^2 x_{34}^2}{x_{13}^2x_{24}^2} \quad \text{and} \quad (1-z) (1-\bar{z}) = \frac{x_{14}^2 x_{23}^2}{x_{13}^2x_{24}^2}
\eeq
The reduced conformal block $\mathcal{F}_\mathcal{O}$ is equal to 
\beq 
z^{S} (z \bar{z})^{\frac{\gamma}{2}} {}_2F_1\left(\ell + \mathbb{S},L-\ell + \mathbb{S}; L+2\mathbb{S};z\right) \,
\eeq
and, in perturbation theory, the three-point function expands as $C_{\mathcal{O}} = C_{\mathcal{O}}^{(0)} + g^2 C_{\mathcal{O}}^{(1)}$, $\mathbb{S} = S + \gamma/2$, and~$\gamma_{\mathcal{O}} =  g^2 \gamma_{\mathcal{O}}^{(1)}$ up to $O(g^4)$.
Then, given a correlator $G(z)$ and the OPE decomposition above, we can read off sum rules from the expansion~\cite{Dolan:2004iy,Plefka:2012rd,Eden:2012rr,Vieira:2013wya}:
\beq
    G(z) = \sum_{S=0}^\infty \left[ (\mathcal{P}_S^{(0,0)}+g^2 \mathcal{P}_S^{(1,1)}) \mathcal{F} + g^2 \mathcal{P}_S^{(1,0)} \partial_\gamma \mathcal{F}\right]\Bigg|_{\gamma=0}
\eeq
with $\mathcal{P}$'s given by simple combinations of the structure constants and anomalous dimensions 
\begin{align}
&\mathcal{P}_S^{(0,0)} = \sum_{\substack{\mathcal{O} \text{ with fixed }S \\ \text{zero-mom}}} \left(C^{(0)}_\mathcal{O}\right)^2 \,, \nn \\
&\mathcal{P}_S^{(1,0)} = \sum_{\substack{\mathcal{O} \text{ with fixed }S \\ \text{zero-mom}}} \left(C_\mathcal{O}^{(0)}\right)^2 \gamma_\mathcal{O}^{(1)} \,,\label{eq:sumrules} \\
&\mathcal{P}_S^{(1,1)} = \sum_{\substack{\mathcal{O} \text{ with fixed }S \\ \text{zero-mom}}} 2\, C_\mathcal{O}^{(0)} C_\mathcal{O}^{(1)} \,. \nn
\end{align}
For example, if we take the correlator for $L=7$, $\ell=2$,
\beq
G(z) = (1+2g^2\mathcal{I}_z) + \frac{2}{1-z}  + \frac{4 - 2 g^2 \mathcal{I}_z}{(1-z)^2} +O(g^4)\,  
\eeq
with the light-cone limit of the box integral given by
\beq
\mathcal{I}_z = \log(1-z) \log(z \bar z)  + 2 \text{Li}_2(z)
\eeq
we obtain the results in \tabref{tab:P-values-L7}:
\begin{placedtable}
    \centering
    \caption{Sum rules for $L=7$, $\ell =2$.}
\label{tab:P-values-L7}
    \fontsize{11}{10}\selectfont
\renewcommand{\arraystretch}{1.45}
    \setlength{\tabcolsep}{8pt}
        \begin{tabular}{c|ccccc}
            \hline
            Spin-$S$ & $0$ & $1$ & $2$ & $3$ & $4$ \\
            \hline
$\mathcal{P}^{(0,0)}$
& $7$
& $0$
& $\frac{11}{4}$
& $-\frac{2}{3}$
& $\frac{28}{39}$ \\
            \hline
$\mathcal{P}^{(1,0)}$
& $0$
& $0$
& $8$
& $-\frac{48}{11}$
& $\frac{54}{13}$ \\
            \hline
$\mathcal{P}^{(1,1)}$
& $0$
& $0$
& $-8$
& $\frac{510}{121}$
& $-\frac{2047}{507}$ \\
            \hline
        \end{tabular}
\end{placedtable}
Using integrability, one can explicitly compute the three-point functions and anomalous dimensions for each zero-momentum state, then perform the sum at each fixed spin $S$ to reproduce such tables \cite{Vieira:2013wya}.

\subsection{Skeleton OPE}
The novel observation here is that we can reproduce each skeleton $G_n$ individually -- with the same conformal block -- by summing over all primaries rather than only zero-momentum ones and by weighting each term by a state-dependent factor $e^{i P n}$.
\beq
G(z) = \sum_{n=0}^{L-1} G_n(z), \qquad G_n(z) = \frac{1}{L}\sum_{\mathcal{O}} \mathcal{F}_\mathcal{O}\times C_{ \mathcal{O}}^2 \times e^{i P_{\mathcal{O}} n}
\eeq

We call each of the terms $G_n$ a \textit{skeleton}; at tree level, each skeleton corresponds to a different tree-level diagram. At one loop, the skeletons correspond to the tree-level diagrams dressed by \textit{hexagons} \cite{Fleury:2016ykk}. To illustrate what the result looks like, we show here the one-loop expressions for the $G_n$ for the choice $L=7$, $\ell=2$.
\beq
G_n=G_n^\text{tree-level}(1+\eta_n g^2\mathcal{I}_z+O(g^4)) \,,
\eeq
where 
\beq
\{\eta_{0},\dots,\eta_{6}\} = \{2,0,-1,0,0,-1,0\} \,.
\eeq
These zeros are well understood both from hexagons \cite{Basso:2015zoa,Fleury:2016ykk,Basso:2015eqa,Bargheer:2018jvq,Coronado:2018ypq} and diagrams \cite{Chicherin:2015edu,Drukker:2008pi}. The basic idea is that skeletons with large bridges suppress loop corrections. For us, however, they are quite non-trivial: Why do some sums weighted by $e^{2iP}$ give a non-zero result but the same sums weighted by $e^{4iP}$ give zero at one loop? (See \tabref{tab:skeleton-ope} for a clear example.)  When seeking a loop generalization of our flow equations, reproducing these zeros should be the zeroth-order test! The next check is to reproduce the $\mathcal{I}_z$'s with the right prefactors $\eta_n$.

\begin{table*}[t]
    \centering
    \caption{   
    Skeleton OPE sum rules for $L=7$, $\ell=2$, and $S=4$. Here $\mathcal{P}_{S,n}^{(a,b)}$ is defined as in \eqref{eq:sumrules} but with sums running over all states (all primaries, of any momentum) \textit{and} each summand dressed by an additional $e^{i P n}/L$ factor. We checked that all these numbers agree with the integrability predictions; as a sanity check, the last column reproduces the data in the last column of \tabref{tab:P-values-L7}.}
    \label{tab:skeleton-ope}
    \renewcommand{\arraystretch}{1.6}
    \setlength{\tabcolsep}{10pt}
    \begin{tabular}{c|ccccccc|c}
        \hline
        & $n=0$ & $n=1$ & $n=2$ & $n=3$ & $n=4$ & $n=5$ & $n=6$ & total \\
        \hline
        $\mathcal{P}^{(0,0)}_{4,n}$
        & $\frac{70}{143}$ & $\frac{19}{429}$ & $\frac{5}{143}$
        & $\frac{5}{143}$ & $\frac{5}{143}$ & $\frac{5}{143}$
        & $\frac{19}{429}$ & $\frac{28}{39}$ \\
        \hline
        $\mathcal{P}^{(1,0)}_{4,n}$
        & $\frac{537}{143}$ & $0$ & $\frac{57}{286}$
        & $0$ & $0$ & $\frac{57}{286}$ & $0$ & $\frac{54}{13}$ \\
        \hline
        $\mathcal{P}^{(1,1)}_{4,n}$
        & $-\frac{891535}{245388}$ & $0$ & $-\frac{33071}{163592}$
        & $0$ & $0$ & $-\frac{33071}{163592}$ & $0$ & $-\frac{2047}{507}$ \\
        \hline
    \end{tabular}
\end{table*}

\section{SoV and OPE coefficients}\label{UntwistedBaxterSoVOPE}

The OPE coefficients for the primaries are given in terms of the untwisted Baxter polynomials $Q^{(0)}(u)$ as follows~\cite{Bercini:2022jxo}\footnote{In this section we are assuming the normalization $Q^{(0)}(i/2) = Q(i/2) = 1$.}:
\beq
\label{untwistedOPEcoef}
\frac{\bigl(C^{(0)}_\mathcal{O}\bigr)^2}{L} = \frac{\Gamma(\ell + S) \Gamma(L-\ell+S)}{\Gamma(L+2S)} \times  \frac{[ Q^{(0)},1]_\ell [ Q^{(0)},1]_{L-\ell}}{[ Q^{(0)},Q^{(0)}]_L} 
\eeq
where the brackets are defined by
\beq
    [ Q_1,Q_2]_l \equiv \int d\nu_{l-1} \prod_{i=1}^{l-1} Q_1(u_i)Q_2(u_i)
\eeq
with the measure factorized as
\beq
    d\nu_l = \frac{1}{l!}\prod_{i=1}^l du_i \ \nu_1(u_i) \prod_{j>i}^{l} \nu_2(u_i,u_j)
\eeq
and simple one- and two-particle factors~\cite{Derkachov:2002tf}:
\begin{align}
\nu_1(u) &= \frac{\pi}{2} \frac{1}{\cosh^2 \pi u}, \\ \nu_2(u,v) &= \frac{\pi (u-v) \sinh \pi (u-v)}{\cosh \pi u \cosh \pi v }\, .
\end{align}
Different solutions of the Baxter equation with length $L$ are orthogonal with respect to this measure:
\beq
[ Q_A^{(0)}, Q_B^{(0)} ]_L \propto \delta_{AB}
\eeq
\subsection{Twisted OPE coefficients}
By introducing a twist in the Baxter equations, there's a slightly different bracket that makes solutions orthogonal~\cite{Cavaglia:2019pow}
\beq
    \langle Q_1,Q_2\rangle_l \equiv \int d\mu_{l} \prod_{i=1}^{l} Q_1(u_i)Q_2(u_i)
\eeq
with the measure defined in \eqref{eq:app-measure}.
The measure factorizes into one- and two-particle factors:
\beq
 d\mu_l=\frac1{l!}\prod_{i=1}^l du_i\,\mu_1(u_i)
 \prod_{j>i}\mu_2(u_i,u_j),
\label{eq:app-measure}
\eeq
where
\begin{align}
\mu_1(u)&=\frac{2\sin\phi\,e^{2\phi u}}{1+e^{2\pi u}},\\
\mu_2(u,v)&=\frac{\pi(u-v)\sinh\pi(u-v)}{\cosh\pi u\cosh\pi v}.
\end{align}
 We can then write the twisted OPE coefficients as 
\beq
\label{twistedOPEcoef}
\frac{C_\mathcal{O}^2}{L} = \frac{\langle Q,1\rangle_\ell \langle Q,1\rangle_{L-\ell}}{\langle Q,Q\rangle_L} 
\eeq
Notice the absence of the combinatorial factors, as compared to~\eqref{untwistedOPEcoef}.\footnote{What is more, we checked that these nice features of the twisted representation also persist at one-loop order.}
\subsection{Untwisting}\label{untwistingAp}
As we untwist, the solutions of the Baxter equation split into two distinct classes: the solutions of the untwisted equations, corresponding to the primaries of SL(2), and the same solutions with extra roots at infinity. In the context of Bethe ansatz/Baxter equations, roots at infinity correspond to symmetry generators -- see e.g. \cite{Faddeev:1996iy, Deguchi_2001,Beisert:2003yb,Beisert:2005di,Beisert:2005fw}. In the untwisted limit, symmetry is restored and solutions with roots at infinity therefore describe the same multiplet as those without roots at infinity. In this setup, they are the $SL(2)$ conformal descendants. In general, for each primary of spin $S$ we have corresponding solutions of degree $S+k$ for all $k\geq1$.

\subsubsection{Primaries}
For primaries, \eqref{twistedOPEcoef} simplifies as we take $\phi$ to zero. The mechanism is that one of the integrals in each of the brackets will be dominated by the $u_1 \rightarrow -\infty$ behavior of the integrand, for example
\beq
\frac{\langle Q,1\rangle_\ell}{[Q^{(0)},1]_\ell}
\sim (2\phi)^\ell\int_{-\infty}^0 du\,e^{2\phi u}u^S(-u)^{\ell-1} = \frac{\Gamma(\ell+S)}{(-2\phi)^S},
\eeq
and similarly for the other brackets.
Putting everything together, the divergent factors cancel and we are left with~\eqref{untwistedOPEcoef}. 
\subsubsection{Descendants}
Upon untwisting, some roots go to infinity and decouple from the finite roots. The $Q$-function of a state with $S$ finite roots and $k$ roots at infinity factorizes into two pieces \cite{Kazakov:2015efa}
\beq
Q(u) \rightarrow Q^{(0)}(u) \times {}_1F_1(-k;L+2S;-2\phi u)
\eeq
Then, 
\beq
\begin{aligned}
\frac{\langle Q,1\rangle_\ell}{[Q^{(0)},1]_\ell}
&\sim(2\phi)^\ell\int_{-\infty}^0 du\,e^{2\phi u}u^S(-u)^{\ell-1}\\
&\qquad\times{}_1F_1(-k;L+2S;-2\phi u).
\end{aligned}
\eeq
Using hypergeometric identities, we get
\beq
\frac{\langle Q,1\rangle_\ell}{[Q^{(0)},1]_\ell}\sim\frac{(-1)^S\Gamma(\ell+S)}{(2\phi)^S}\times\frac{(L-\ell+S)_k}{(L+2S)_k}
\eeq
For the denominator, the relevant integral coincides with the norm of the Laguerre polynomials and we have:
\beq
\frac{\langle Q,Q\rangle_L}{[Q^{(0)},Q^{(0)}]_L}\sim\frac{\Gamma(L+2S)}{(2\phi)^{2S}}\times\frac{k!}{(L+2S)_k}
\eeq
For the rest of the integrals we can simply replace $Q$ by $Q^{(0)}$. We therefore find that the three-point function for the $k$-th descendant of a primary $\mathcal{O}$ is given by:
\beq
C_{\mathcal{O},k}^2 = \frac{(\ell + S)_k (L-\ell + S)_k}{k!(L+2S)_k} \times \bigl(C^{(0)}_\mathcal{O}\bigr)^2
\eeq

Finally, summing over all descendants of the primary $\mathcal{O}$, we get its total contribution to the OPE (at tree level)
\beq
\frac{\mathcal{F}_{\mathcal{O}}}{z^S}\!=\! \sum_{k=0}^\infty z^k \frac{C^2_{\mathcal{O},k}}{{(C_{\mathcal{O}}^{(0)})^2}} \!=\! {}_2F_1(\ell+S,L-\ell+S;L+2S;z)
\eeq

\section{\texorpdfstring{$L=1$}{L=1} intuition} \label{L1kernelAp}

In this section we explore in detail the simplest $L=1$ case where everything can be computed exactly in closed form from multiple points of view. While not necessary, we find all these manipulations very useful to develop intuition for the general case. 

\subsection{Computing the kernel by brute force}

Writing $\tau=e^{i\phi}$, the Baxter polynomials normalized to $\langle Q_S,Q_S\rangle_1=1$ read
\beq
Q_S(u)=\tau^S\,{}_2F_1\left(-S,\tfrac12+iu;1;1-\tau^{-2}\right) \,.
\eeq
Using the standard integral representation of the hypergeometric function, we can write 
\beq
\!\!\! Q_S(u)=\frac{\cosh\pi u}{\pi}\int_0^1 dt\,
t^{-1/2+iu}(1-t)^{-1/2-iu} q(t)^S
\label{eq:L1Q}
\eeq
where $q(t)\equiv \tau(1-t)+\tau^{-1}t $. 
Substituting this representation into the sum \eqref{defkernel} we can trivially perform the geometric sum over spin to obtain
\begin{multline}
K^{(1)}(u,v)=\frac{\cosh\pi u\cosh\pi v}{\pi^2}\\
\times\int_0^1\!dt\int_0^1\!ds\,
\frac{t^{-\frac12+iu}(1-t)^{-\frac12-iu}s^{-\frac12+iv}(1-s)^{-\frac12-iv}}
{1-zq(t)q(s)}.
\end{multline}
The remaining two integrals can now be done, leading to the following compact representation for the $L=1$ kernel:
\begin{multline}
K^{(1)}(u,v)=\frac1{1-z}\left(\frac{1-z\tau^2}{1-z}\right)^{i(u+v)}\\
\times{}_2F_1\left(\tfrac12+iu,\tfrac12+iv;1;\chi\right),
\label{eq:app-L1-compact}
\end{multline}
where $\chi=-4z\sin^2\phi/(1-z)^2$.

\subsection{Completeness: the limit \texorpdfstring{$z\to1$}{z → 1}}
Using the well-known hypergeometric connection formula, we can rewrite the kernel \eqref{eq:app-L1-compact} as 
\begin{multline}
K^{(1)}(u,v)=\\
\frac{\Gamma(-i(u-v))}{\Gamma(\frac12-iu)\Gamma(\frac12+iv)}
\frac{(1-z)^{i(u-v)}}{(1-z\tau^{-2})^{\frac12+iu}(1-z\tau^2)^{\frac12-iv}} \\
\times{}_2F_1\Big(\tfrac12+iu,\tfrac12-iv;1+iu-iv;\frac{1}{1-\chi}\Big)
+(u\leftrightarrow v).
\end{multline}
This representation is the most convenient one to study the $z\to 1$ behavior since in this limit the hypergeometric goes to $1$, leaving its prefactor plus its $u\leftrightarrow v $ symmetrization. 

The interesting terms are those in the numerator. As $z$ approaches $1$, the $(1-z)^{i(u-v)}$ factor oscillates rapidly in $u-v$ forcing $v\simeq u$ as a distribution. Close to this locus, the gamma function behaves as a simple pole and the terms in the denominator basically reconstruct the (inverse) one-particle measure. All in all, as a distribution, we find, upon symmetrization in $u\leftrightarrow v $, that as $z\to 1$, the~$L=1$ kernel becomes \footnote{See also \cite{Derkachov:2021wja}.}
\beq
\frac{i}{2\pi}\lim_{z\to 1} \frac{1}{\mu_1(u)} \frac{(1-z)^{i (u-v)}-(1-z)^{-i (u-v)}}{u-v} = \frac{\delta(u-v)}{\mu_1(u)}  \,,
\label{eq:L1delta}
\eeq
or, equivalently,
\beq
\lim_{z\to 1}\int d\mu_1(v)K^{(1)}(u,v)f(v)=f(u).
\eeq
Of course, this is precisely what we ought to get. Indeed, if we plug $z=1$ in the definition \eqref{defkernel} we get precisely the completeness relation
\beq
\lim_{z\rightarrow 1}K^{(1)} \equiv \sum_S  \frac{Q_S(u)Q_S(v)}{\<Q_S,Q_S\>_1} = \frac{1}{\mu_1(u)} \delta(u-v) \, 
\eeq
as advocated in the introduction. 
This exercise illuminated in which incarnation this delta function shows up: It is the sinc one. 

\subsection{The Flow Equation(s)}
Of course, for higher $L$ we cannot compute the kernel by brute force, so the flow equations come in handy. In this section we illustrate how to use them to recover the $L=1$ result written at the beginning of the section. The best way to do that is to combine the first equation in \eqref{eq:Dequation2} and covariance \eqref{eq:covariance} to get one set of reduced flow equations
\beq
\chi \partial_\chi H_\chi = D_u H_\chi = D_v H_\chi \, .
\eeq
Plugging in an ansatz for $H_\chi$ as a power series $H_\chi = \sum_n h_n \chi^n$ we find, explicitly,
\begin{align}
n h_n(u,v) &= i(u-i/2)(h_n(u-i,v)-h_n(u,v))  \\
&= i(v-i/2)(h_n(u,v-i)-h_n(u,v))  
\, .
\end{align}
The solution to these equations is fixed up to an $i$-periodic function $c_n(u,v)$:
\beq
h_n = c_n(u,v) (1/2+iu)_n (1/2+iv)_n
\label{eq:hn}
\eeq
Combining now the ansatz and the second equation in \eqref{eq:Dequation2}, we have the reduced equations
\begin{align}
\chi (1-\chi) \partial_\chi H_\chi &= (\bar{D}_u + \chi (1/2+iv) )H_\chi \\ 
&= (\bar{D}_v + \chi (1/2+iu) )H_\chi\, .
\end{align}
Plugging \eqref{eq:hn} into this expression gives the coefficients $c_n(u,v)$ in terms of a single function $c_0(u,v)$. Finally, using that $K=1$ for $z=0$, we fix $c_0=1$ and, therefore,
\beq
h_n = \frac{(1/2+iu)_n (1/2+iv)_n}{(n!)^2}
\eeq
which implies 
\beq
H_\chi(u,v) = {}_2F_1(1/2+iu,1/2+iv;1;\chi)
\eeq
Combining this result with the prefactor, we reproduce the kernel derived in the previous section.
\section{Mellin and the \texorpdfstring{$L=2$}{L=2} kernel} \la{L2kernelAp}
An alternative way to solve the flow equation is to go to a Mellin representation,
\begin{equation}
\mathcal M_x[f](u)\equiv\frac{\cosh\pi u}{\pi}
\int_0^1 dx\,x^{-\frac12+iu}(1-x)^{-\frac12-iu}f(x).
\label{eq:mellindef}
\end{equation}
In that language, shifts in $u$ become multiplication by $x$ and multiplication by $u$ becomes derivatives in $x$. As a result, the $L=1$ flow equations are two simple PDEs, one for each $j$:
\begin{equation}
z\partial_z\widehat K(x_1,x_2)
=\left(x_j+\frac{\tau^2}{1-\tau^2}\right)
\partial_{x_j}\widehat K(x_1,x_2).
\end{equation}
This equation can be solved by the method of characteristics: equations for $j=1$/$j=2$ imply the Mellin kernel only depends on $z$ through $z q(x_1)$/$z q(x_2)$, hence the solution should be a function of $z q(x_1) q(x_2)$, with $q$ defined just below \eqref{eq:L1Q}:
\beq
\widehat K(x_1,x_2)=f\!\left(zq(x_1)q(x_2)\right)
\eeq
The boundary condition for this equation can be derived from the transform of the $\delta$-function from the previous section $\eqref{eq:L1delta}$ and it reads
\begin{align}
\lim_{z\rightarrow 1} \widehat K &= \frac{1}{2 i \sin \phi}\frac{1}{\tau^{-1}x_1 x_2-\tau (1-x_1)(1-x_2)}\\
&= \frac{1}{1-q(x_1)q(x_2)}.
\end{align}
It follows then from the method of characteristics:
\begin{align}
    \widehat K(x_1,x_2) = \frac{1}{1-z q(x_1)q(x_2)}.
\end{align}
\subsection{Covariance and \texorpdfstring{$L=2$}{L=2}}
For higher $L$ it's convenient to use the covariance form \eqref{eq:covariance}, which translates as
\beq
\widehat K(\vec x)
=\frac{\widehat H_\chi
\bigl(X_z(x_1),\ldots,X_z(x_{2L})\bigr)}
{(1-z\tau^2)^L\prod_{j=1}^{2L}D_z(x_j)}
\eeq
with the factors $D_z$ given by
\beq
D_z(x)=1-(1-\lambda_z^{-1})x,\qquad\lambda_z=\frac{1-z\tau^2}{1-z},
\eeq
and the variables changed to 
\beq
X_z(x)=\frac{x}{\lambda_zD_z(x)}
\eeq
In that language, the $L=1$ case discussed before reads
\beq
\widehat H_\chi(x_1,x_2)=\frac{1}{1-\chi x_1x_2}
\eeq
By writing down an expansion for $\hat{H}_\chi$ around $\chi=0$ we can fix its coefficients -- in the same way we did for the $L=1$ result in the previous section -- to obtain the closed formula as the reciprocal square root of polynomials in the variables $X_z(x)$:
\beq
\widehat H_\chi(\vec x)=
\frac{1}{\sqrt{\bigl[1-\chi(e_3-2e_4)+\chi^2e_4\bigr]^2
-4\chi^2e_4\bar e_4}}
\eeq
where $e$ and $\bar{e}$ are the symmetric polynomials in $X_z(x_j)$ and $1-X_z(x_j)$. More precisely, $e_3 = \sum_{i<j<k}\allowbreak X_z(x_i)\allowbreak X_z(x_j)\allowbreak X_z(x_k)$, $e_4 = \prod_{i=1}^4 X_z(x_i)$ and $\bar{e}_4 = \prod_{i=1}^4 (1-X_z(x_i))$. In principle one could use that strategy to find the $\chi$ expansion of the kernel for any length $L$ but, as we saw in the main text, we don't need the closed result in order to compute the correlators we are interested in.
\subsection{Integration and evaluation in Mellin space}
Another useful feature of this representation is that integration against the one-particle measure $d\mu_1(u)$ corresponds to evaluating the corresponding variable $x$ at the point $x_\phi = \frac{\tau^2}{\tau^2 - 1}$:
\beq
\int d\mu_1(u) \mathcal{M}[f](u) = f(x_\phi)
\eeq
To see that we just need to integrate both sides of \eqref{eq:mellindef}: the measure kills the $\cosh(\pi u)$ in the prefactor and leaves only exponentials in $u$. The integral over $u$ then yields a $\delta$-function that performs said identification.

In particular, integration of the $L=2$ kernel against two $1$-particle measures gives
\begin{multline}
\int d\mu_1(v_1) d\mu_1(v_2) K(u_1,u_2,v_1,v_2)= \\ 
\mathcal{M}\left[\frac{1}{\sqrt{1-2z(x_1+x_2-2x_1x_2)+z^2(x_1-x_2)^2}}\right].
\label{eq:intkernelmellin}
\end{multline}
As expected, all dependence on the twist has disappeared and the remaining integrated kernel has now a simple Mellin representation. Transforming the result back to $u$-space gives the beautiful result
\begin{multline}
\int d\mu_1(v_1) d\mu_1(v_2) K(u_1,u_2,v_1,v_2)= \\ 
{}_2F_1\!\left(\tfrac12+iu_1,\tfrac12-iu_2;1;z\right) {}_2F_1\!\left(\tfrac12-iu_1,\tfrac12+iu_2;1;z\right).
\end{multline}
From that it's easy to evaluate $G_0$ and $G_1$ by simply evaluating the formula above for $(u_1,u_2)=(i/2,i/2)$ and $(u_1,u_2)=(i/2,-i/2)$, respectively. That reduces the hypergeometric functions to simple functions such as $1$ and $1/(1-z)$, thus matching the results in the main text:
\begin{align}
    G_0 &= {}_2F_1(1,0;1;z) {}_2F_1(0,1;1;z) = 1 \\
    G_1 &= {}_2F_1(0,0;1;z) {}_2F_1(1,1;1;z) = \frac{1}{1-z}
\end{align}

Indeed, even if we didn't know how to perform the transform to get the hypergeometric function, we could still evaluate \eqref{eq:intkernelmellin} at $u=+i/2$ and $u=-i/2$. The way it works is that in Mellin space, these simply map to evaluations at $x=0$ and $x=1$, respectively. 
\begin{align}
    \mathcal{M}[f](i/2) = f(0)\,, \qquad  \mathcal{M}[f](-i/2) = f(1)\,.
\end{align}
As such, $G_0$ is computed from \eqref{eq:intkernelmellin} by simply setting $x_1=x_2=0$ there and $G_1$ is computed by setting $x_1=0$ and $x_2=1$. Again, that reproduces the expected result.

\section{Covariance}
\label{covarianceAp}
In this section we will prove some of the formulas in the main text related to the structure of the Baxter operator and its cousins. 
As discussed there, the main ingredient in most of the proofs is the algebra that the pieces of these operators satisfy. One of the relevant commutation relations, repeated here for convenience, is 
\beq
[D_I,\bar{D}_I]= D_I-\bar{D}_I
\eeq
Beyond that, there are two other relations between the operators $D$ and $\bar{D}$ and the sum of rapidities $C$,
\begin{align}
[D_{I},C_{I}] &= -i D_I + C_I - i \frac{L}{2} \label{eq:DCcommutator} \\
[\bar{D}_{I},C_{I}] &= i \bar{D}_I + C_I + i \frac{L}{2} 
\label{eq:DbarCcommutator}
\end{align}
Now, to prove covariance, we remember that the two equations \eqref{eq:Dequation2} must be compatible, since the operators labeled by $I$ and $\bar{I}$ by definition act on different sets of variables. That is, $K$ needs to satisfy
\beq
[D_I,\bar{D}_{\bar{I}}] K(I,\bar{I}) = 0
\eeq
Although the commutator is identically zero, the equation is nontrivial because acting on $K$ with $\bar{D}_{\bar{I}}$ generates a term with $C_I$, which doesn't commute with $D_I$. Then, using \eqref{eq:DCcommutator} and \eqref{eq:DbarCcommutator} we arrive at the desired result
\begin{multline}
\left[(1+z)\partial_\phi-2z(1-z)\cot\phi\,\partial_z\right.\\
\left.{}+2z(C_I+C_{\bar{I}}+L\cot\phi)\right]K=0.
\label{eq:covariancepde}
\end{multline}
This differential equation implies the covariance of the kernel, equation \eqref{eq:covariance}. To see that, notice that after extracting the prefactor described in the main text, the equation simplifies as follows:
\beq
\left[(1+z)\partial_\phi-2z(1-z)\cot\phi\,\partial_z\right]H_{z,\phi}=0.
\eeq
Since the differential operator above kills $\chi$, any function of $\chi$ \eqref{eq:chi} satisfies the condition above and we must have $H_{z,\phi} = H_\chi$.

\section{Twist independence}
\label{PhiIndAp}
In \secref{IntSec} we claimed the integrated kernel is independent of the twist $\phi$ and in this section we will prove that. First, to simplify notation, let $V=(v_{1},\ldots,v_L)$. Then the integrated kernel is given by the following expression
\beq
\mathcal{J}_z(V) = \int d\mu_\ell(U) d\mu_{L-\ell}(W) K(U,W,V)
\eeq
The first step in proving twist independence here is to note that the product of the two measures in the integrand can be written as a single measure of size $L$ divided by a product of two-particle factors
\beq
B_\ell(I) \equiv \frac{d\mu_\ell(U)d\mu_{L-\ell}(W)}{d\mu_L(U,W) } = \binom{L}{\ell} \prod_{u \in U, w \in W} \frac{1}{\mu_2(u,w)} \, ,
\eeq
where $B_\ell$ does not depend on the twist and we define $I=U\cup W$. This function behaves nicely under shifts of rapidities. In particular, as we will show at the end of this section, it satisfies a simple identity involving the difference operators $D$ and $\bar{D}$:
\beq
(D_I-\bar{D}_I) B_\ell(I) = 0
\label{eq:Bidentity}
\eeq
With this ingredient we now proceed to show the twist independence of the integrated kernel. By differentiating $\mathcal{J}_z$ with respect to $\phi$, the derivative acts either on the measures or on the kernel since, importantly, $B_\ell$ is twist independent:
\beq
\partial_\phi \mathcal{J}_z(V) = \int d\mu_L(I) B_\ell(I) (\partial_\phi + 2 C_I + L \cot\phi) K(I,V)
\eeq
With some algebra, it is possible to use the equations \eqref{eq:Dequation2} and \eqref{eq:covariancepde} to eliminate $\partial_\phi$ and write the RHS in the convenient form
\beq
\partial_\phi \mathcal{J}_z(V) = \frac{i(1-z)}{2\sin^2\phi} \int d\mu_L(I) B_\ell (D_I-\bar{D}_I)^t K(I,V) = 0
\eeq
with the superscript $t$ indicating transposition with respect to the measure $d\mu_L(I)$, which corresponds to shifting contours. The second equality follows then from the nice relation \eqref{eq:Bidentity} satisfied by $B_\ell$. To prove it, let's look at the seemingly unrelated quantity first,
\beq
R = \text{Res} \left(\frac{i (t-i/2)^L}{\displaystyle\prod_{u\in U}(t-u)\prod_{w\in W}(t-w-i)} + (U\leftrightarrow W)\right)\,,
\label{eq:residuesR}
\eeq
where $\text{Res}$ denotes the sum of residues at the finite poles $u$, $u+i$, $w$ and $w+i$. The residues at $u_j$ and $w_j$, combined, are simple and read
\beq
\frac{1}{B_\ell(U,W)}\sum_{x \in I} \frac{i (x-i/2)^L e^{-i \partial_x} B_\ell(U,W)}{\prod_{y\neq x} (x-y)} 
\eeq
and similarly for $u_j+i$ and $w_j+i$. Therefore, the sum of all these residues can be written simply as 
\beq
R = \frac{1}{B_\ell(U,W)} (D_I - \bar{D}_I + 2 i C_I)B_\ell(U,W)
\eeq
On the other hand, we can compute $R$ by computing the residue at infinity. Then, from the large $t$ expansion of the expression inside the parentheses in \eqref{eq:residuesR},
\beq
R = 2 i C_I
\eeq
By comparing with the previous equation we arrive at the nice identity \eqref{eq:Bidentity}.

\section{The other case} \label{OtherCase}

In this appendix we explain how to compute the skeleton $G_n$ for $n<\ell$ (the case $n>L-\ell$ follows a similar derivation). To do that, we need to take the kernel $K$, set $n$ rapidities to $-i/2$, $L-n$ rapidities to $i/2$, and then integrate it against measures $\mu_\ell$ and $\mu_{L-\ell}$. To simplify notation, write
\beq
F=K\bigl(U,W,(+i/2)^{L-n},(-i/2)^n\bigr),
\label{eq:Fz}
\eeq
with the powers $L-n$ and $n$ denoting repeated arguments. As in \eqref{eq:Gn}, $n$ counts the number of rapidities set to $-i/2$. We write its integral as 
\beq
G_n=\int d\mu_\ell(U)\,d\mu_{L-\ell}(W)\,F\,.
\eeq 
For the case at hand, $L-n$ is greater than both $\ell$ and $L-\ell$, allowing us to build the two different partitions: $I_1=U\cup\{(+i/2)^{L-\ell}\}$ and $I_2=W\cup\{(+i/2)^\ell\}$. We can then sum the first equation in \eqref{eq:Dequation2} for each of the choices, and use covariance \eqref{eq:covariancepde} to eliminate the sums of rapidities $C$. Finally, using the reduction formula \eqref{eq:reduction} to set the appropriate rapidities to $i/2$, we get
\beq
(D_U^{(\ell)}+D_W^{(L-\ell)})F
=e^{i \phi}\sin\phi\left[\partial_\phi-2iz\partial_z+\frac{2inz}{1-z}\right]F.
\eeq
As for the case discussed in the main text, this equation only contains operators with shifts of the two sets of variables $U$ and $W$. This allows us now to integrate both sides by moving the $D$'s and $\partial_\phi$ past the measures and get a differential equation for the skeleton $G_n$
\beq
\left[(1-z)\partial_z-n\right]G_n(z)=0,
\eeq
where in the above we have used that the integrated kernel is independent of the twist (see \appref{PhiIndAp}). With the boundary condition $G_n(0)=1$, we find:
\beq
G_n=(1-z)^{-n}
\eeq

For $L-\ell<n< L$, a similar argument applies. We therefore reproduce the expected answer \eqref{sumSkeletons}.

\section{Properties of the kernel} 
\label{properties}
In the main text we established that it's possible to uplift the equations for the kernel to equations for its integrals. However, the kernel itself is an interesting object. Let's list  some of its properties. We are adding a subscript $z$ to the kernel in this appendix since some of these properties involve transformations of the cross-ratio.
\paragraph{Cyclicity.} From the twisted conservation of momentum $e^{i P L} = e^{-2i\phi S}$,
\beq
K_z((i/2)^L,X) = K_{z \tau^2}((-i/2)^L,X)
\eeq
\paragraph{Composition.} As a direct consequence of orthogonality,
\beq
\int d\mu_L(Y) K_{z_1}(X,Y) K_{z_2}(Y,Z) = K_{z_1 z_2}(X,Z)
\eeq
In particular, when $z_2 = 0$, the identity reduces to the next property:
\paragraph{Projection.}
\beq
\int d\mu_L(Y) K_{z}(X,Y) = 1
\label{eq:projection}
\eeq

\paragraph{Decoupling.}
Pairs of $+i/2$ and $-i/2$ decouple from the kernel:
\beq
K_z(+i/2,-i/2,X)=\frac{K_z^{(L-1)}(X)}{1-z},
\label{eq:decoupling}
\eeq

The decoupling property \eqref{eq:decoupling} is very remarkable when viewed in terms of the kernel's definition! It relates sums of $Q$-functions in a spin-chain of length $L$ to sums of $Q$-functions on a spin-chain of length $L-1$. 
To show just how much information it captures, let us rederive the first case of \secref{IntSec}. Since $n \geq \ell$, we can use the identity $\ell$ times on equation \eqref{eq:Fz}:
\beq
F = \frac{1}{(1-z)^\ell}K_z^{(L-\ell)}(U,W,(+i/2)^{L-n-\ell},(-i/2)^{n-\ell}) .
\eeq
But now the integrals against the measures follow immediately from the projection property \eqref{eq:projection} and we recover the desired result. Using the reduction \eqref{eq:reduction}, it can be shown that both sides of equation \eqref{eq:decoupling} satisfy the same equations.

\section{Casimir operator and untwisted flow} \label{untwistedAp}

As we saw, twisting the $Q$-functions and the Baxter operators simplifies the problem in many ways: it drastically reduces the conformal block to $z^S$ and it provides a simple spin operator $\hat{S}$ that allows us to represent the kernel as an evolution of the identity operator. Nevertheless, it's still possible to find a differential equation for the untwisted kernel. Let's start by defining this object. It is a function of $2L-2$ variables $u_1,\dots,u_{L-2},v_1,\dots,v_L$:
\beq
K^{(0)} = \sum_{\mathcal{O}} \mathcal{F}^{(0)}_{\mathcal{O}} \, \frac{Q(u_1)\dots Q(u_{L-2}) Q(v_1)\dots Q(v_L)}{[Q,Q]_{L}}\, ,
\label{untwistedKernel}
\eeq
where the sum runs over all primaries $\mathcal{O}$ and with the untwisted conformal block\footnote{This is the usual conformal block dressed with the $\Gamma$-functions so that the summand of \eqref{untwistedKernel} only has $Q$-functions.} $\mathcal{F}^{(0)}$ given by, dropping the subscript,
\beq
\mathcal{F}^{(0)}=B(\ell+S,L-\ell+S)\,z^S\,{}_2F_1(\ell+S,L-\ell+S;L+2S;z),
\eeq
where $B$ denotes the Euler beta function. On one hand, all the dependence on the cross-ratio $z$ is inside the conformal block, which satisfies
\beq
\mathcal{D}_z\mathcal{F}^{(0)}=S(S+L-1)\,\mathcal{F}^{(0)}
\eeq
with the Casimir differential operator in $z$ realized by
\beq
\mathcal{D}_z=z^2(1-z)\partial_z^2+z\left[L-(L+1)z\right]\partial_z-\ell(L-\ell)\,z
\eeq
On the other hand, the untwisted $Q$-functions satisfy the Baxter equation
\beq
\hat{\mathcal{B}}_u \circ Q(u) =T(u) Q(u) \label{baxteruntwisted}
\eeq
where the Baxter operator is given by \eqref{baxterop} with the twist $\phi$ set to zero and the transfer matrix is now given by
\beq
T(u)=2u^L-\left[S(S+L-1)+\tfrac{L(L-1)}{4}\right]u^{L-2}+\sum_{k=0}^{L-3}c_k u^k
\eeq
In a similar fashion to the twisted case worked out in the main text, we can use the Baxter operator to construct an operator that acts on a product of $Q$-functions and returns the Casimir eigenvalue:
\beq
\mathcal{C}_{\vec w}\prod_{j=1}^{L-1}Q(w_j)=S(S+L-1)\prod_{j=1}^{L-1}Q(w_j)
\eeq
Explicitly, this Casimir operator is given in terms of shifts in the rapidities as
\beq
\mathcal{C}_{\vec w}=\sum_{j=1}^{L-1}\frac{(w_j+\tfrac i2)^L\bigl(e^{i\partial_{w_j}}-1\bigr)+(w_j-\tfrac i2)^L\bigl(e^{-i\partial_{w_j}}-1\bigr)}{-\prod_{k\neq j}(w_j-w_k)}.
\eeq
We therefore have the untwisted flow equation
\beq
\left(\mathcal{D}_z-\mathcal{C}_I\right)K^{(0)}(I,\bar I)=0
\eeq
for any subset $I$ of $L-1$ of the $2L-2$ rapidities, with $\bar I$ its complement.

\section{Finite coupling details}
\label{app:finite}

The finite-coupling three-point function with one SL$(2)$ operator and two BPS ones factorizes into a normalization piece and the resummation of adjacent and bottom magnons~\cite{Basso:2022nny,Basso:2025mca}. When the non-BPS state has twist $L=2$ and the BPS states have twist $p$, we denote these factors by $\mathcal{A}$ and $\mathcal{B}_p$. In \cite{Ekhammar:2026sli}, the authors conjecture the formula \eqref{B2} for the Konishi operator, with spin $S=2$. We further conjecture the formula holds for any spin, which is supported by tests up to five loops for spins $S=2,4,6,8$. In this appendix, we will show how this formula can be compactly rewritten with the introduction of a twist and how it recovers the weak-coupling SoV formulas with and without twist. In all of this section we are using the conventions of \cite{Ekhammar:2026sli}.

\subsection{Introducing a twist}

The claim is that the adjacent part $\mathcal{A}$ for the configuration we are dealing with is simply given by $\mathcal{B}_0$, which takes the form
\beq
\mathcal{B}_p = \det(\mathbf{1}_{m\leq n} + H^{(p)}_{m n})
\eeq
with the kernel $H$ defined as
\beq
H^{(p)}_{m n} = \frac{2i}{g} \sum_{a=1}^\infty \int \frac{du}{2\pi}  \chi^{i j} \mathbf{q}_{2m-2,i}^{[+a]} \widetilde{\mathbf{q}}_{2n-1,j}^{[-a]}
\label{Hmn}
\eeq
 For $p=0$, the integral in the $m=n=1$ element of the matrix is divergent. Indeed, if we introduce $z=u+ia/2$ then when both $u$ and $a$ are large, the integrand behaves as 
 \beq
 F(u,a) = \frac{g^3 \hat{\alpha}}{z \bar{z}^2}(z^{M_2} \bar{z}^{M_1} - z^{M_1} \bar{z}^{M_2})
 \eeq
 with $\{M_1,M_2\} = \frac{1}{2}\{\Delta-S,\Delta+S-2\}$ and this clearly only converges if $\Delta<2$ which is not the case 
since $\Delta = 2+S+\gamma$ and both $S$ and $\gamma$ are bigger than zero. To remedy that we insert a twist inside the kernel in order to regulate this divergence~\footnote{In the mirror sums of \cite{Basso:2022nny} that should amount to twisting the mirror magnons by exponentials (e.g. $e^{2i\varphi u}$), but we will not discuss that here.}. Consider the deformed kernel
 \beq
 \widetilde{H}^{(p)}_{m n} =  \frac{2i}{g} \sum_{a=1}^\infty \int \frac{du}{2\pi} \, \textcolor{blue}{\cos(2 \varphi u)} \, \chi^{i j} \mathbf{q}_{2m-2,i}^{[+a]} \widetilde{\mathbf{q}}_{2n-1,j}^{[-a]} \, .
 \label{Hmntwisted}
 \eeq
 As $\varphi \rightarrow 0$, the important piece of the integral is the $u,a \rightarrow \infty$ region, similar to what we discussed in \appref{untwistingAp} and to the analysis performed in~\cite{Basso:2019diw}. In that case, the sum over $a$ becomes another integral and, using that the integrand is invariant under $z\rightarrow -z$,
 \begin{align}
 \widetilde{H}_{11}^{(0)} &\sim i \frac{g^2 \hat{\alpha}}{\pi} \int_{\mathbb{C}} \frac{d^2 z}{z \bar{z}^2} \textcolor{blue}{e^{2 i \varphi \text{Re}(z)}}\left(z^{M_2} \bar{z}^{M_1} - z^{M_1} \bar{z}^{M_2}\right)  \\ 
 &\sim\textcolor{blue}{\varphi^{2-\Delta}}i^{S+1} g^2 \hat{\alpha} \left(\frac{\Gamma(M_2)}{\Gamma(2-M_1)} + \frac{\Gamma(M_2-1)}{\Gamma(1-M_1)}\right)
 \\ 
 &\sim\textcolor{blue}{\varphi^{2-\Delta}} \times  g^2 |\hat{\alpha}|(M_2-M_1)\frac{\Gamma(M_2-1)}{\Gamma(2-M_1)}
 \end{align}
For sufficiently small $S>0$~\footnote{The first equality in the equation below is only true for $2<\Delta<6$ since in general more terms would be divergent, but the representation is useful to better understand the origin of the relative factors between $\mathcal{A}$ and $\mathcal{B}_2$.}, this is the leading divergent piece and therefore we can extract it from the determinant, leaving the minor with indices $m,n\geq 2$ and twist sent to zero. Since $H$ satisfies $H^{(p)}_{m+1,n+1} = H^{(p+2)}_{m,n}$, the determinant of that minor is simply $\mathcal{B}_2$. Restoring the $\Delta$ and $S$ dependencies, we then get the desired relation
\beq
\lim_{\varphi \rightarrow 0} \varphi^{\Delta-2} \widetilde{\mathcal{B}}_0 = g^2 |\hat{\alpha}|(S-1)\frac{\Gamma((\Delta+S-4)/2)}{\Gamma((4-\Delta+S)/2)} \times \mathcal{B}_2 = \mathcal{A} \nn
\eeq

\subsection{From Fredholm determinant to SoV}
The bilinear combination in $\mathbf{q}$'s in equation \eqref{Hmn} is nothing but the $(4|4)$ transfer matrix $\mathbf{t}_{a,1}$ (see \cite{Basso:2022nny,Ekhammar:2026sli}). On the other hand, at weak coupling (more generally whenever wrapping contributions can be dropped) those are proportional to the $(2|2)$ transfer matrices $T_a$ as~\cite{Basso:2015eqa},
\beq
\mathbf{t}_{a,1}(u) \rightarrow (-1)^a (x^{[+a]}x^{[-a]})^{L/2}\frac{T_a(u^\gamma)}{\prod_{j=1}^S h_{a,1}(u^\gamma,u_j)} \, .
\eeq
At leading order, that combination takes a very simple form involving the Baxter polynomial $Q$ for the state in consideration~\cite{Basso:2015eqa}. Using this expression, the $m=n=1$ element of \eqref{Hmntwisted} with $p=0$ becomes simply
\beq
\widetilde{H}_{1 1}^{(0)} = \sum_{a=1}^\infty \int \frac{du}{i \pi} \sum_{r,s=\pm} r s \frac{ Q^{[r a+s]}(u)}{Q(\frac{i}{2})\, (u-\frac{ia}{2})} \textcolor{blue}{\cos(2 \varphi u)}
\eeq
Take $\varphi>0$ so we can close the integration contour in the UHP and pick the pole at $u=i a/2$, which yields
\beq
\widetilde{H}_{11}^{(0)} = \sum_{a=1}^\infty  \frac{Q^{[2a]}(\frac{i}{2})-Q^{[2a-2]}(\frac{i}{2})-Q(\frac{i}{2})+Q(-\frac{i}{2})}{Q(\frac{i}{2})} \textcolor{blue}{e^{-\varphi a}} \nn
\eeq
Using the zero-momentum condition and collecting the coefficients of each $Q$-function, we can write the result as an integral with the SoV measure. Indeed, after the identification $\varphi = -2 i \phi$ and using the normalization $Q(i/2) = 1$, we have exactly
\begin{align}
1+\widetilde{H}_{11}^{(0)} &= (1-e^{-\varphi})\sum_{n=0}^\infty \textcolor{blue}{e^{2 i \phi n}} \frac{Q^{[2n]}(\frac{i}{2})}{Q(\frac{i}{2})} \,.
\end{align}
But this is nothing but the evaluation of 
\beq
\frac{1}{Q(\frac{i}{2})}\int du\, \mu_1(u) \, Q(u) = \langle Q,1\rangle_1
\eeq
if we evaluate this integral by residues, from the poles of $\mu_1$, see (\ref{mu1})!

In this appendix, the twist $\varphi$ was merely an artifact to regulate the integrals and had no physical meaning nor affected the $Q$-functions, as opposed to what we did in the main text. Some people call this \textit{fake twist} $\varphi$ while the main text one would be \textit{real twist} $\phi$. If we end up taking them to zero in the end, both are equivalent but the second one usually leads to much nicer finite twist expressions. It would be a very interesting question to find the analogue of the real twisted $Q$-system and SoV representations for twisted states with non-zero momentum so we could isolate various skeletons in a QSC based approach. 

\raggedright
\bibliography{references}

\end{document}